# Introducing a Research Program

# for Quantum Humanities – Applications

Author 1 (corresponding author) Astrid Bötticher, Friedrich-Schiller University of Jena,Bachstraße 18k 07743 Jena, Research Fellow Innsbruck Quantum Ethics Lab (IQEL) Universität Innsbruck, Innrain 15, 6020 Innsbruck

Author 2 Zeki C. Seskir, Karlsruhe Institute of Technology - Institute for Technology Assessment and Systems Analysis (ITAS), Karlstraße 11 76133 Karlsruhe

Author 3 Johannes Ruhland, Friedrich-Schiller University of Jena, Carl-Zeiß-Straße 3 07743 Jena

**Abstract**

Quantum computing is a rapidly developing field in the second wave of quantum development, with the potential to revolutionize a wide range of industries and fields of study. As the capabilities of quantum computers continue to advance, they have the potential to significantly impact society and the way we live, work, and think. This makes it important for scholars from a variety of disciplines to come together and consider the implications of these technologies. How this was done has already been explained and published in an abstract way in a joint research paper. But how exactly these abstract theoretical approaches come into an implementation could not be shown so far. The present article shows exactly this.

**Introduction**

Quantum humanities is an interdisciplinary field that brings together scholars from the humanities, social sciences, and the arts to explore the intersection between quantum technology and these disciplines. By examining the impact of quantum computing on society and considering the theoretical and philosophical implications of these technologies, quantum humanities aim to provide a more comprehensive understanding of the potential consequences of these advancements. Alongside this is an attempt to transform and advance the humanities using quantum computing. Through a new type of computing and questions adapted to it, the quantum computer also has the potential to change the humanities. There are many applications of quantum computing that have the potential to significantly impact the humanities and social sciences. For example, quantum computing could be used to improve the efficiency and accuracy of financial modeling and investment analysis, to optimize the design and operation of renewable energy systems, to improve logistics and supply chain management, to design and synthesize new materials and drugs, and to analyze and interpret medical data. Additionally, quantum computing could be used to improve the accuracy and efficiency of language processing tasks, such as translation and summarization. As the capabilities of quantum computing continue to advance, it is important for scholars from a variety of disciplines of social science and humanities to come together and consider the implications of these technologies for society and the way we live. Through interdisciplinary research, we can better understand the potential consequences of these advancements and identify the questions and issues that need to be addressed in order to ensure that they are used ethically and responsibly.

This article presents a literature review that gives an insight to the most actual research within quantum humanities, structured according to the core elements of quantum humanities. The applications are presented among research topics that have a great impact on humanities and social sciences as such and that can depict the transformation to quantum humanities. These applications are centered around the topics: 1. Finance, 2. Security, 3. Energy, 4. High-Performance Computing/Scientific Computing, 5. Transportation, 6. Health, 7.Natural Language Processing. The analysis of the applications is followed by the question of their significance. What implications do the leap innovations associated with quantum computing hold? What role do they play in the organization of our everyday life, our society, and the norms in which we move? In part, our considerations here go into the realm of the speculative. This is about identifying a space of possibilities that the innovations can potentially have. In a third step, the innovations that accompany technical innovation and social change are examined for their significance for the humanities and social sciences. In a final step, we would like to at least briefly outline the theoretical Implications that accompany our findings and make them available for further discussion, even though this is only possible in the form of mental indents.

1. **A Research Program for Quantum Humanities**

    a. **A holistic qualitative research program**

The field of quantum humanities is not simply digital humanities, as it focuses specifically on the impact of quantum computers and the implications of their unique properties, such as indeterminacy, on the way we approach and solve problems. There are many potential applications for quantum computing in the fields of humanities and social sciences. Some examples of these applications include using quantum algorithms to analyze large datasets, such as those found in social media or demographics, or using quantum computers to model and understand complex systems, such as those found in economics or political science. Other fields that may be impacted by quantum computing include linguistics, archaeology, and history, as these fields often involve the analysis of large amounts of data and the search for patterns within that data. To sustainably improve these fields through the use of quantum computing, it will be important to consider not just the technical aspects of the technology, but also the broader implications and impacts on society. Additionally, it will be important to consider the ethical and social implications of using quantum computing in these fields and ensure that the technology is used responsibly and transparently.

For a comprehensive research program, we identified four qualitative fields of interest. These four fields are central to working through the change caused by quantum technology around us and in our fields of work. Quantum engineering, rather than quantum computing, is the research object of Quantum Humanities. But in the paper written here, quantum computing alone will be the subject. This is simply due to the scope. Quantum engineering and its various fields, such as quantum sensing, are again so comprehensive that the topic could not be covered "in passing" in one article. Also, it will not be possible to address all four fields, because the research area is so comprehensive that it would probably require a multi-volume publication to show them in their entirety. Instead, we will content ourselves with examining applications of quantum computing, and only those where, in a preliminary wk for this article, we have also seen actual changes.

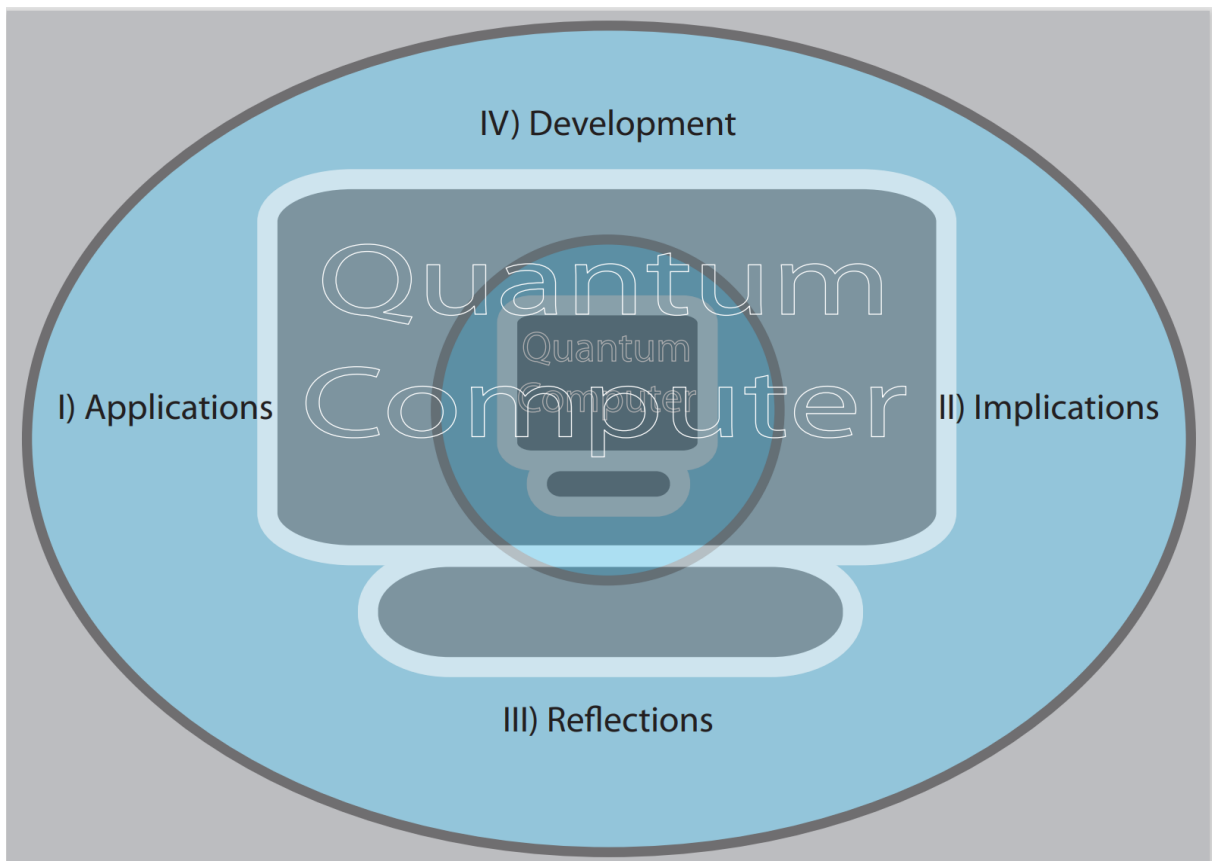

Fig. 1: Core elements of Quantum Humanities (Designer: Gerhard Kiegerl).

We have developed a holistic vision of research and we have defined four main fields of interest for quantum humanities and regard them as the research field. We understand, in contrast to digital humanities (that is mainly a field of research that is concerned with the application of computational tools and methods to humanities), that quantum humanities is a broad research field and is not solely composed by the use of a quantum computer. These fields of research are comprised of: i) *application* of quantum computing for addressing questions from humanities and social science research or research that can have an impact on implications or reflections; (ii) *reflection* of the methods, techniques, and impact; and (iii) societal, cultural and social *implications* that are expected as a potential for upheaval; and (iv) looking at how development processes and ecology are structured and how developments will be driven forward with what emphasis. This results in a hybrid field of research, located between humanities, social sciences, and quantum computing. Each individual segment of the research suggests a whole series of investigations that need to take place in order to take a holistic view of the transformation phenomenon.

## 2 Applications

The applications we want to discuss today are not from the Humanities but are applications from other fields, but which could have a direct impact on the Humanities. We are convinced that Quantum Humanities is not just about the applications of quantum computing in the humanities, but that it is a comprehensive research program in which various developments have a changing influence and this can be investigated and demonstrated in the context of qualitative research. But even today's applications, which are not at all specifically concerned with quantum-based advancement of the humanities, are transforming social science and the humanities. They change them because they change society in questions of everyday action or, for example, in that new institutions must be developed or institutions change. Today, as part of the qualitative research program we are developing, we are demonstrating how applications can be studied and made fruitful for probing new fields in social science and the humanities by examining developments in quantum computing. In doing so, we chose cases where we believed that the advancements in the field in question were striking. For the selection of cases, it was not crucial whether there was an obvious impact on social science and humanities or not, rather (and this is also related to the fact that quantum computing is not yet mature today), the crucial question was whether quantum computing could be used to achieve changes in the respective field that was unambiguous.

The application of quantum computing is being successfully advanced in various fields right now and Honeywell names five of them (Honeywell 2020). In this paper, we will focus on applications that help us to reflect generally on technique and implications (different from this approach is Miranda 2022). We will not be able to describe here how quantum gates are calculated or how a loop is built in. This makes more sense for an investigation to be described specifically, but not for an overview like this one. Some areas that we assume will have a major impact on social sciences and humanities are presented here in a more systematic way. Doing so enables us to name the most important developments for the humanities and social sector, and on the other hand, we can get a first idea of the nature of the phenomenon of quantum technology in the context of social science and humanities theory, so that we can develop, in dashes, a first research program for which the Quantum Humanities Network stands. In the end, the importance for humanities and social sciences is not only explained by feedback to our theories or methods, i.e. to our tools, but also by the fact that a field of interest arises for investigation. We must state here that we are far from being able to take all innovations and developments into account. Nevertheless, an impression should arise here for possible topics of investigation as well as for the application of new techniques and new tools for our sciences themselves.

### 2.1.1 Security

Quantum security (Harlow/Hayden 2013) is a central topic of security policy, which has increasingly addressed the concept of digital sovereignty in the context of (post)digitization (Capgemenini 2022). In the field of technology, there is a desire for quantum sovereignty in hardware and software on the one hand, and an urgent need for security to safeguard central societal systems such as the military, the intelligence service, the financial sector, and critical infrastructure like transport systems on the other (Wimmer/Moraes 2022).

Modern encryption technology is based on an asymmetry: it is very fast to multiply two prime numbers, but it takes an incogitable long time to decompose a very large whole number into its prime components (Bennett/Brassard 1984). The Shor algorithm (Shor 1997) to factorize whole numbers is used for prime number decomposition and is a quantum algorithm. RSA encryption (Rivest–Shamir–Adleman public-key cryptosystem) is based on this. For today's systems, RSA encryption is the gold standard, but previous work shows that the RSA standard, which is used to encrypt corporate secrets or sensitive government communications, for example, as well as bank transfers and other sensitive areas of communications, will not be able to withstand quantum computing (Gidney/Ekerå 2021). This is why the quantum computer poses a risk to the digital systems developed today (Vlachou/Rodriguez/Mateus/Paunković 2015). Therefore, for questions of security, the standardization of quantum computer-resistant alternatives in particular is at the forefront of activities and migrational activities are the main activities of governments (NSA News 2020).

Risk analysis against quantum computers distinguishes different processes that interact with each other and contribute to the security assessment. These include (among others) the time in which existing systems must securely protect data stores or the time to migrate existing systems into a post-digital age (Mosca/Piani 2019). For migration purposes (among others) a distinction is made between post-quantum cryptography and quantum cryptography (BSI 2022). In 2016 NIST proposed a competition (NIST 2022) to develop quantum-safe cryptography and was looking for proposals for algorithms to proceed with migration processes for existing systems (Computer Security Resource Center 2022) or quantum cryptography, there is more effort is needed, as great technological infrastructural systems (like optical fiber deployment and satellite networks[1]) need to be built (Chen/Zhang/Chen 2021) and today, we witness a geopolitically motivated race for the establishment of quantum supremacy, that is coined (among other) with the concept of crypto agility (Johnson et al 2019; Giles 2019).[2] Another important development for security studies might be the transfer of social network monitoring analysis to quantum computing for terrorism detection purposes (Zahedinejad/Crawford/Adolphs 2019) or other community monitoring applications relevant to security (Bisconti/Corallo/De Maggio et al 2009). Since a very large amount of data often has to be collected here in real-time, and since this data must be determined as precisely as possible - also to protect civil rights - it could be an immense improvement to apply quantum computing to this type of use (Grover 1996).

**2.1.2 Finance**

The financial industry can, with some overlap, be divided into three branches: banking, (commodity) trading on financial markets (including investment banking), and insurance. Market prediction and the management of associated risks, trade optimization, and target group identification are among the core success factors in each of them. As one of the most global nimble industries, where milliseconds can make the difference between profit and loss, and that is operating on a quantitative, monetary basis naturally, it comes as no surprise that research into quantitative methods of evaluation has had a strong position in most companies worldwide. Exploring the potential of Quantum Computing from a theoretical and prototypical perspective three core fields have been identified over the last years:

- Monte Carlo simulations, especially with the intent to evaluate the risk exposure of an investment portfolio
- Portfolio Optimization (under the inevitable market risk)
- Methods of Machine Learning applied to target group identification and prediction. (Albaretti 2022/ Bobler 2021)

---

[1] There are several governmental activities to develop a quantum satellite network. Peoples Republic of China that already has a developed satellite network, EU has developed a new satellite and the USA also prove efforts to develop large scale infrastructure. See DARPA: Quantum Key Distribution Network (darpa.mil)https://www.darpa.mil/about-us/timeline/quantum-key-distribution-network. See also: ESA - Secure communication via quantum cryptography: https://www.esa.int/Applications/Telecommunications_Integrated_Applications/Secure_communication_via_quantum_cryptography. See also: China Builds the World's First Integrated Quantum Communication Network (scitechdaily.com) https://scitechdaily.com/china-builds-the-worlds-first-integrated-quantum-communication-network/

[2]. See also national activities in Europe: https://www.qunet-initiative.de/ie QuNET-Initiative - QuNET.

Simulation is popular within the whole finance industry (Swayne 2022) The intrinsically uncertain value of complicated options or financial derivatives, e.g., cannot be derived analytically but must be determined from a simulation based on possible development paths of underlying asset values. Also, the quantification of risk exposure of a company as a whole is most often done through estimation of the long, highly improbable tail of the profit distribution and calculating indicators from it ("Value at Risk", "Conditional Value at risk …"), a procedure also mandated by many supervising authorities. For large investment portfolios, this may take hours to run, with QC reducing it to minutes; theory gives an asymptotic quadratic performance boost (Egger et al 2020).

Improvements in the optimization of large portfolios, and for a bunch of ML methods are still more dramatic (Orùs/Mugel/Lizaso 2019), enabling real-time applications that operate in sync with market fluctuations. The instability of qubits in contemporary hardware propagates to intrinsic unreliability of results, especially in large-scale applications, often limiting their use to Late Prototypes. Contrary to accounting, 100% precision is often not mandated in the applications listed above; current research in the field of Noisy Intermediate-Scale Quantum Calculations (NISQ) is expected to give a timely performance boost (but Bouland et. al.2020) taking a critical position). For the same reason, further use cases in the industry are expected to show up in an unexpected and unsteady manner in the next years to come.

**2.1.3 Energy**

There are several ongoing and predicted application areas of quantum computers for the energy sector, such as materials research on batteries (Mitsubishi 2019) and fuel cells (DLR 2021), optimizing power (Exxon Mobil 2019), and energy grids (Ajagekar/You 2019), simulation and reliability analyses (Giani et al 2021) and expanding reservoir production(Parney/Garcia/Womack 2019). Furthermore, there are expectations of quantum computers vastly outperforming supercomputers when it comes to energy efficiency (Villalonga 2020,) and studies on roughly estimating when this might happen are being conducted (Jaschke/Montangero 2022). This will be very important for those experts, studying climate change and transformational processes. Companies like Mitsubishi Chemical, ExxonMobil, Gazprom Neft, and BP have been actively experimenting with one or more of these application areas. Therefore, in the energy sector, the use of quantum computing has the potential to be widely adopted if advantages over classical supercomputers can be shown.

**2.1.4 High-Performance Computing (HPC)**

High-performance computing is a major topic for quantum computing and Möller and Vuik name some applications connected to quantum computing, like green Aircraft, flood prediction, and realistic models of protein interactions within human bodies (Möller/Vuik 2017). A joint study by Atos and IQM (Perini/Ciarletta 2021) found that "*...76% of HPC data centers worldwide plan to use quantum computing by 2023, and that 71% plan to move to on-premises quantum computing by 2026*"; the concept of HPC units with quantum processing units (QPU) has been in circulation since at least 2017 (Britt/Mohiyaddin 2017). This relies on two main arguments: First, quantum computers (or QPUs) themselves are expected to require considerable computational support systems, especially for error correction which is necessary for fault-tolerant quantum computation. This makes an HPC data center a suitable environment for a quantum computer.

Second, it is a long-known fact that quantum computers only allow speed-ups for algorithms that can utilize certain mathematical structures. A very obvious example is multiplication. The multiplication process itself does not get any speed-up from quantum computers, prime factorization (which is practically the reverse of multiplication) gets an exponential speed-up, hence enabling Shor's algorithm. This effectively creates two subsets of problems, ones that make sense to use quantum computers for and the ones that don't. However, complex algorithms with commercial value usually do not fit neatly tintoone of these subsets. Certain parts of the algorithms can get substantial speed-ups from quantum computers while the rest don't. In such cases, a hybrid approach that utilizes different processing units for different purposes is usually adopted. A similar example of this is the division of labor between graphics processing units (GPUs), central processing units (CPUs), and recently tensor processing units (TPUs). Therefore, adding QPUs to this set of processing units seems like a logical next step for HPC data centers.

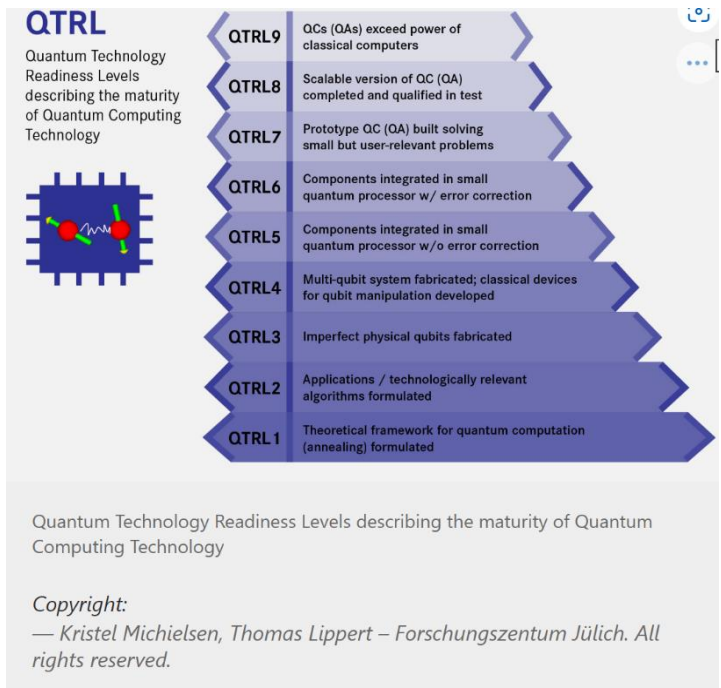

There are already several cloud service providers experimenting with integrating quantum devices into their systems. IBM Quantum, Amazon Bracket, Microsoft Azure, and D-Wave's Hybrid Solvers are some early examples. This is expected to effectively equip HPC centers that employ QPUs with enhanced capabilities which in return will enable them to outperform their fully classical counterparts. Also, [3]However, the current technology readiness levels of the quantum devices, sometimes referred to as QTRLs, developed by Michielsen & Lippert (2022), are not high enough for such a migration to make sense for competitive purposes. Some of the applications are myths (Singh Jattana 2020). Analogous to other assessments of technology maturity, such as those made by NASA, a scale for the development status of the Quantum Computer was developed at Forschungszentrum Jülich (Michielsen/Lippert 2022).[4] It is intended to track the development of Quantum Computing and develop a general understanding of the stage of development applications are in.

Quantum Technology Readiness Levels describing the maturity of Quantum Computing Technology

*Copyright:*
— Kristel Michielsen, Thomas Lippert – Forschungszentrum Jülich. All rights reserved.

**2.1.5 Transportation**

There have been studies and efforts in adapting quantum computers for advantage in four of the major areas of transportation; airlines (IBM Institute for Business Value 2022a), maritime logistics (IBM Institute for Business Value 2022a), railroads (Saran 2021), and the automotive industry (BMW Group Challenge 2022). Companies like Boeing, Airbus, BMW, Daimler, and Deutsche Bahn have been experimenting with quantum-enhanced solutions, and even live testing for public transportation was implemented in 2019 with a collaboration between Volkswagen AG and the city of Lisbon for a pilot project to provide quantum computing-based traffic optimization (Yarkoni et al 2020).

---

[3] There has been research going on to connect HPC with QC, yet the path is still long. In Jülich research Centre, there exists a project to develop a "High Performance Computer – Quantum Simulator hybrid" (High Performance Computer – Quantum Simulator hybrid | HPCQS). There is also research paper dealing with the infrastructure of High performance Computing and Quantum Computing in Europe that deals with the topic of a connection between HPC and QC: Binosi, D., Calarco, T., Colin de Verdière, G. (2022): European Quantum Computing & Simulation Infrastructure. Microsoft Word - 20220202_HPC-QCS-JWP-final.docx (qt.eu).

[4] See also: "The Technology Readiness Level of Quantum Computing Technology (QTRL) is based on NASA's assessment of space technologies. The quantum computer scale ranges from level 1 (formulation of a theoretical framework) to level 9 (quantum computers exceed the power of conventional computers). Currently, Michielsen considers development to be at level 5: components have been integrated into a small quantum processor without error correction." Kristel Michielsen (fz-juelich.de) https://www.fz-juelich.de/en/research/information-and-the-brain/quantum-technologies/kristel-michielsen.

Most of the proposals in these different areas rely on three assumptions. First, quantum computers are expected to be better for optimization. This has been a selling point since the early days of commercial quantum computing and annealing with quadratic unconstrained binary optimization (QUBO) algorithms. The development and popularization of quantum approximate optimization algorithms (QAOA) in the mid-2010s (Farhi/Goldstone/Gutmann 2014) transformed this to an almost mainstream assumption that quantum computers are useful for optimization tasks. Second, due to the exponential nature of the Hilbert space, large scale quantum computers are accepted as promising avenues for big data analytics, given that there can be efficient fitting of the data to the devices. This has been an active area of research for the last decade (Wiebe/Braun/Lloyd 2012), and due to the relatively small sizes of the current devices, the jury is still out on whether this is practically useful or feasibly implementable. Finally, it is assumed that there are some overlaps between industrially relevant problems in transportation and the problems that rely on mathematical structures which benefit from exponential speed-ups from quantum computers. There are already at least three groups (Arute/Arya/Babbush 2019; Madsen/Laudenbach/Askarani 2022; Zong/Wang/Deng 2020) around the globe (in the US, China, and Canada respectively) that demonstrated their quantum computational devices perform a task (albeit, practically useless task) faster by several orders of magnitude. The following question for the transportation industry is, whether any relevant problems for the sector can benefit from these exponential speed-ups.

Combination of these three assumptions, which are, quantum computers are suited for optimization purposes, can handle big data effectively, and can provide exponential speed-ups, makes these devices really intriguing for many applications related to transportation. Traffic data (whether it is cars, planes, trains or ships) comes in big volumes and in real-time, and performing optimization algorithms on it requires considerable computational power. This means, an optimization algorithm that takes too long to yield results is not very useful. However, if quantum algorithms (even if they are less accurate than their classical counterparts) can run much faster and respond to the changes of conditions in real-time, they would be adopted fast and wide in these industries.

**2.1.7 Health**

A report published by the IBM Institute for Business Value in 2020 titled "Exploring quantum computing use cases for healthcare" puts forward three use cases for quantum computers in the healthcare industry; (i) diagnostic assistance, (ii) precision medicine, and (iii) pricing. An early example of how quantum algorithms might be used for diagnostic assistance is a collaboration (Case Western Reserve University n.d.) between Microsoft and Case Western Reserve University in 2018, where they reported 30% more precise findings, and up to three times faster scans for Magnetic Resonance Fingerprinting (MRF) using quantum-inspired algorithms that work on existing machines.

Similarly, approaches like quantum network medicine (Maniscalco/Borrelli/Cavalcanti 2022) aim to explore efficient search mechanisms for new drugs and new drug combinations using quantum algorithms. Research on the molecular drug development that might later be utilized for personalized medicine has been a selling point to get pharmaceutical companies involved in quantum computing. However, current quantum computers are able to effectively simulate molecules such as $H_2O$ at best (Eddins/Motta/Gujarati 2022), while pharmaceutically relevant molecules and interactions of them with other types of molecules require much further capabilities than what is available today.

Finally, the pricing aspect is presented as an application of risk analysis and insurance. The concept of quantum risk analysis (Woerner/Egger 2019) has been around for some time, and there have been several applications of quantum computers for finance (Orùs/Mugel/Lizaso 2019); subsection 2.1.2). Therefore, adapting these applications to risk assessment for insurance purposes and pricing of health insurance is another potential application of quantum computers in the healthcare industry. Changes in health care and the impact of quantum computing on public health systems is not foreseeable today. However, the ability to develop medicine tailored to the patient or to better detect and target tumor cells is a potentially system-changing opportunity for the increasingly rampant cancer epidemic in the developed world (Abbott 2021; Solenov et al 2018; Fraunhofer 2021). This gives rise to potentially important tasks for the sociology of medicine or questions of the sociology of technology, but also for political science and law and ethics.

**2.1.8 Natural Language Processing**

Natural Language Processing (NLP) is a method that is relevant for all humanities and social sciences, since it is used as a quantitative technique for example for the investigation of semantic networks, sentiment analysis or social network analysis and other techniques for the investigation of social or humanistic questions. Useful for quantitative linguistics, quantitative law, social science statistics and relations analysis in the different social science fields. The central problem of these methods is usually a very large amount of data and the emergence of complex developments and processes that are difficult to calculate with the existing standards of computing or take a very long time and are then hardly interpretable (for example, the famous snowball problem of social science network analysis). For some of these problems the Grover algorithm is useful, yet the research community to further explore QNLP is small (Villalpando et al 2021).

Quantum Natural Language Processing (QNLP) is a concept first appearing in the literature in 2016 (Zeng & Coecke) and the first conference on QNLP was organized in 2019[5]. The concept is emerging in the literature, however slowly (Abbaszade et al 2021) A quick literature search[6] yields 16 results for arXiv, 7 for Scopus, and 3 for Web of Science. There exists a high level Python library (Kartsaklis, et al., 2021) accessible via Github (Cambridge Quantum's QNLP team 2021) for those aiming to play around and implement QNLP algorithms. One interesting relation of QNLP is with music composition, which is also highlighted in a conference on the topic titled "1st International Symposium on Quantum Computing and Musical Creativity'' organized in 2021 (Interdisciplinary Centre for Computer Music Research 2021). Presentations on this topic started appearing in other conferences as well (such as QSD 2022 (QWorld a 2022; QWorld b 2022)). Exploration of these emerging topics and their potential applications in the field of social sciences and humanities may be a worthy avenue of research. Quantum natural language processing is used for example for machine translation, sentiment analysis, relationship extraction, word sense disambiguation and automatic summary generation (NEASQ b. n.d.).

---

[5] There have been several conferences held in Oxford and Cambridge for example:
http://www.cs.ox.ac.uk/QNLP2019/ and QNLP 2022 (cambridgequantum.com)
[6] Query: ("Quantum Natural Language Processing" OR QNLP); Date: 21/08/2022

## 2.4 Implication

We have outlined the impact of the introduction of quantum computing for individual science sections. Now we would like to briefly discuss how the quantum computer will have a transformative impact on many areas of social life. We have pointed out developments that we assume will trigger the most important transformation processes, affecting the humanities and social sciences in the near future. These transformations will be described among the lines of societal, cultural and social *implications.* The use of quantum computers has important implications for the realm of societal organization as vast infrastructure must be secured and replaced to be quantum safe. These implications are centered around the topics: 1. Finance, 2. Security, 3. Energy, 4. High Performance Computing/Scientific Computing, 5. Transportation, 6. Health, 7. Natural Language Processing. We believe that the quantum computing research discussed here in the application section requires a societal debate to examine and evaluate the impact of the technology on science, industry, people, and society. Hence, we discuss the implications of quantum computing research, though with the limit that the era of quantum computing has just begun and that it is impossible to see the great inventions of the future.

### 2.4.1 Finance

Deployment of QC applications will no doubt accelerate financial markets that already today are operating on a split second, global basis. Through their trading success, companies already having QC facilities at hand will push this trend even further, even without pursuing an explicit "competing in time" strategy. Automated trading through bots will increase as algorithms become both faster and more reliable. Under unusual market conditions without precedent and the ability of Artificial Intelligence Algorithms to learn, this provokes the risk of detrimental, unforeseen and chaotic behavior, should several algorithms competing in one market run awry simultaneously.

Quantum computing in the cloud democratizes access and is available to those smaller companies that cannot afford their own hardware. Quantum supremacy on the hardware level will be hard to achieve on a company level, chances are higher on a national and especially supra-national level. Political actors may try to limit access to "their" cloud facilities in order to profit from technological advances

To reap the benefits mentioned, new and more powerful algorithms need to be developed and must be cast into Quantum ready software. Specialists are expected to remain rare (Hughes et al 2022) "Programmers" in the conventional sense will have to acquire new skills in programming. Making use of the unique properties of QC requires understanding of Quantum Physics and its fundamental ramifications - qualities that even only a part of professionally trained and experienced physicists typically possess (Rietz 2022). In addition, to reap the most of the benefits from QC, new algorithms extending and replacing existing approaches need to be developed, with some being at the verge of commercial application already (Kerenidis/Prakash 2017; Bouland 2020; Egger et al 2020; Montanaro 2020). Companies or nations that can establish a close collaboration between algorithm designers, programmers and hardware specialists will gain much larger benefits than those merely transferring existing approaches to a quantum cloud platform (Vijayaraghavan V 2021). Power will shift towards larger firms and conglomerates that can bring in their economies of scale and have experience in Innovation Management and with QC prototypes. Ironically, many of them have already hired trained physicists to create and maintain sophisticated models of capital markets.

Governments could and probably will tighten their compliance control over the financial industry, as daily reporting (and subsequent first level conformance checking with advanced Artificial Intelligence methods) could in principle replace current Quarterly Reports. This pressure could be propagated down, putting comparable requirements upon companies to which core credit exposures or securities engagement exist. Supranational regulations such as Basel III and IV show the way.

Even with the advent of QC, one of the core issues of the financial industry, i.e. building an early warning system for imminent global financial crisis probably may stay too ambitious a goal (as may be the question of reactions to once in a lifetime market constellations) and here QC may create a false sense of security. One of the few, relatively popular approaches developed by Orùs (Orùs/Mugel/Lizaso 2019) and later Ding (Ding/Lamata et al 2019) tries to model a great number of financial entities and their nonlinear interactions to analyze alarm signals and imminent crisis developing and spreading. While being amenable to run fast on Quantum hardware, it remains unclear if this paradigm can model real life markets reliably and free of modeling or computational artifacts. In this important case, the present QC approaches seem much like fitting the problem to the tool. Likewise, in the closely related field of anomaly detection and the creation of resilient organizations, it is conceptual work rather than computing power that is the limiting factor.

### 2.4.2 Security

In terms of development and use, the quantum computer and its capability are woven into a web of (international) political relationships already set by digitization (Paul 2007), and the post-digital technology is thus given a specific political meaning (de Wolf 2017) and is dependent on institutions (North 2019) for its use and utility. Within the realm of security, the quantum computer is understood as a technology that poses a systemic risk to economies (Mosca/Piani 2019), and could trigger catastrophic cascading effects if used for conflict-oriented political purposes (Ekert/Renner 2014), and an urgent need for action on polycentric regulation (World Economic Forum n.d.) is indicated. This is especially important regarding the topic of blind computing, that becomes something like the new TOR in a much more fundamental sense, if you like (of course this analogy has central weakness) (Broadbent/Fitzsimoni/Kashefi 2009). In addition to being a gamechanger on the international geopolitical stage, quantum computing is a gamechanger for issues related to the security and stability of entire systems or their institutions and infrastructures. At the same time, the application of quantum computing to specific security contexts, such as the detection of terrorist entities in real time or the analysis of security-related data in disaster management, can become an important new tool for threat analysis. In addition to these more traditional security applications for law enforcement and intelligence or the military, there are important advances coming for disaster response that should not be overlooked. These include the management of floods and also other issues related to global warming that pose a challenge to disaster management systems - not least new weather models.

### 2.4.3 Energy

One relevant risk posed by quantum computers is enabling cyber attacks that were previously not possible, rendering previously safe systems vulnerable. There are of course efforts being undertaken to prevent such attacks via transitioning these systems to either post-quantum cryptography algorithms or to quantum cryptography schemes. However, it is a race between quantum computers being developed and the critical cyber infrastructure being upgraded. Considering the timeline of the quantum threat (Mosca/Piani 2019), there is a chance that certain critical infrastructure, especially in regions with less available resources to spend on upgrading their existing systems, might be vulnerable to these kinds of attacks. Therefore, working on prevention and mitigation scenarios that don't rely on overhauling entire infrastructures is needed.

### 2.4.4 High Performance Computing

HPC resources are also tools beyond their operational uses, they are frequently used as a show of force, where countries and companies demonstrate their computational capabilities. A good example of this is the TOP500 (Top 500 n.d.) list that ranks and details the 500 most powerful non-distributed computer systems in the world. A reflection of this can be seen in the race for quantum computers as well. In 2019 Google announced it has reached 'quantum supremacy' (Arute/Arya/Babush 2019), where the US presidents daughter Ivanka Trump (senior advisor to the Trump administration) announced "The US has achieved quantum supremacy!" on Twitter (Trump 2019). The following year a group of researchers at USTC in China announced they also achieved quantum computational advantage. Although there have been calls from researchers to keep quantum computing global and open (Biamonte/Dorozhkin/ Zacharov 2019), recent developments (such as the invasion of Ukraine) makes it practically not possible for countries to consider quantum computers as mere tools of scientific research.

### 2.4.5 Transportation

Societally more relevant applications of quantum computing for the transportation sector are expected to be for autonomous vehicles and traffic management systems in the near to mid term. Although it may seem like a purely technical area at first, the transition period presents an opportunity with strong social implications, as it might be used for re-imagining and designing public spaces. As concepts such as carfree city centers get adopted around the globe, optimal distribution of limited resources for public transportation gains importance. Furthermore, cities are living organisms, which frequently encounter disruptions in their regular mode of public transportation and traffic flow. As discussed previously, quantum computers might hold the potential to shorten simulation times significantly, which can enable city administrations to respond and redirect public transportation resources in-time. However, such capabilities require retraining of the employees and development of necessary infrastructure is a time and resource heavy endeavor. Indicating that before cities decide to adopt such solutions to the transportation and mobility related issues, there needs to be serious consideration, giving rise to the necessity of exploration and research into the topic.

There are many other similarly relevant applications in the transportation sector for quantum computers, such as intercity travel, national and international flight route management, maritime traffic, and so on. An additional layer to these discussions is whose concerns (and parameters) are going to be prioritized during the process. As the real-time optimization capabilities increase in a steady manner, there will be choices between different parameters on prioritizing traffic noise over commute time, and more particularly, prioritizing of different elements within the transportation system (a recent example is Germany's decision to prioritize cargo trains with coal over passenger trains (Dezem 2022). All in all, quantum computers present an opportunity for further involvement of the public into transportation affairs and require debates on how to distribute the limited resources of logistics and mobility systems.

## 2.4.7 Health

Health is a fundamental part of human life and healthcare systems have been an essential element of any modern functioning society. With the advent of digitalization and artificial intelligence, these systems are expected to undergo severe changes. For example, in recent years there have been an increasing number of studies on whether A.I. systems can outperform doctors on diagnostics, which they sometimes do (McKinney/Sieniek/Godbole 2020). However, such discussions were held in 1970s and 1980s with the 'expert systems' approach to A.I. as well, and after much research, those models failed to deliver the expected performance. This time around, A.I. researchers and cloud computing companies (such as IBM with their Watson Health) are more confident in their models. Simultaneously, all these companies are also working on developing quantum computers. This might result in several societally relevant outcomes.

First, quantum computers, especially the ones that rely on superconducting qubits, are centrally operated machines. This means, they are operated at dedicated data centers (as discussed in the HPC subsection). As health data and diagnostics become integrated into these cloud computing based business models, sharing of data with these companies needs to become an industry norm. Although there are certain proposals, like blind computing (Broadbent/Fitzsimoni/Kashefi 2009) to circumvent this necessity, they require additional steps and constraints to be introduced, effectively raising both the cost and the level of expertise required to implement quantum enabled diagnostics systems. As measures such as GDPR and sectoral regulations for data management (especially in finance and health) are being developed, in parallel, there are technical systems being developed that require annulment of the protections brought forth by these regulations. Similar to the Chatcontrol (Breyer, n.d.; European Commission 2022) discussions, the proper amount of exceptions in data protection for access to automatized healthcare services might become a societally relevant issue as health applications that run on quantum computers gets further developed.

Second, as healthcare systems further get integrated with the computational way of providing services, there will be a need for new linkages to be formed and further skills to be blended into the existing networks. For example, ongoing discussions in the field of algorithmic decision-making in healthcare (Grote/Berens 2020) will only be exacerbated as methods developed for fair machine learning, such as explainability, can't be used for quantum fair machine learning models (Elija Perrier 2021). New methods of integrating non-medical technical expertise into healthcare systems, how to assess the fairness of ongoing operations, and how to effectively distribute resources for optimal outcomes are topics for social debate where the existence of quantum computers might have some role in.

### 2.4.8 Natural Language Processing

Natural Language Processing, that uses sentences as networks, is already widely used in the digital industry, in security and science (especially linguistics). Whether for advertising or the use by authorities (police, intelligence, military) or scientific usage such as surveys and conflict research - data analytics and its network applications today are mainly based on the idea that sentences are networks and their respective tools. The use of the quantum computer as a tool will help to undertake analysis of greater data volume. It might also be helpful for heterogeneous data to be analyzed in conjunction. As more precise results are achieved with quantum computers, a new field of legal science could emerge here for example (of course, only if the relevant algorithms are developed) that makes use of modern technology (Floridi/Sanders 2001). But there are no use cases today that we know of, though there are some practical works presented connected with QNLP (Rudolph/Bashige/Touissant/Katabarwa 2022).

## 2.5 Reflection

*In this section, we focus on the reflection* of the methods, techniques and impact that quantum computing will have on our monocultural sciences and how it will gradually change the things we do and how we do them.We have now pointed out individual uses and, in a further step, named transformative processes that will also affect the humanities and social science. Now we would like to link the applications and the associated transformation processes back to the scientific task and briefly outline the ways in which the humanities and social sciences will be affected. Here, however, it must be said restrictively that the effects on the specific sciences have not yet been systematically considered, and we can only give broad strokes of thought here (Vermaas 2017). We are so far at the beginning that we first have to sound out which innovations could have any significance at all - in order to then be considered systematically for the individual humanities and social science strands in a second step. Of course, this task is far too complex here - already using the example of political science, we can briefly outline how specialized the individual sciences are: In addition to the study of war and violence (and their termination), there are bureaucratic processes that are so highly differentiated that, for example, pension issues have to be considered separately from questions about long-term care insurance or social welfare. Nevertheless, all these specializations are united by the fact that they depend on the processing of huge amounts of complex data (while changes can occur quickly) in order to be able to make any statements at all.

### 2.5.1 Finance

After the commercial introduction of mainframes in the 1960s, PCs and Networks in the 1980, and Smartphones and Mobile Computing towards the Millennium, QC carries the potential to spark a 4th IT revolution. Yet several traits will set this revolution apart. Foremost, it carries all characteristics of Restrained Innovation. This phenomenon is typically described through the advent of jet airplanes towards the end of WWII. All necessary technology innovations had been in place several years before, with the notable exception of highly temperature resistant metal alloys for the manufacture of turbine blades. This innovation been achieved, the introduction of jet airplanes progressed with dazzling speed. In the QC hardware case, the restraining factor is the high number of sufficiently (though not necessarily 100%) reliable Qubits.[7] Once solved, the pace of adoption will be faster than expected by an unexperienced observer, and market players that have taken preparation through conceptual development and prototypes will be able to reap their benefits. Most industries involved operate in high speed markets for their own business (finance, life science, Intelligence agencies …) and thus are used to a fast pace of innovation, pushing speed still further. Adoption will no doubt immediately take place on a global scale, as markets and all of the major players are operating globally. This speed may come as a complete surprise to many stakeholders in economy, politics and society. As empirical study by one of the authors (JR) shows, an overwhelming majority of Small and Mid-sized Enterprises (SMEs) tends to defer all relevant action concerned with QC to vaguely defined "other actors" (government, cloud providers, OEMs, …) – if they have heard of QC at all. Ironically, many of them are completely unaware, that practices such as "store now, decrypt later"[8] are affecting their business already today. Some small niche innovators aside, it is unlikely that smaller companies will be able to shape the QC development path. What they need to have though, is fast reaction, high resilience and alertness to innovations well beyond their current level. Taking the perspective of Industrial Economics, QC adoption will start with industries, where speed and mobility is a key success factor, be in the product development, the production of or reaction to market fluctuations. Large firms and conglomerates will leverage their growing expertise with QC to change the rules of the marketplace, and may even profit from operating their own, cutting edge hardware.

From a legal perspective, fast moving industries and Restrained Innovation pose serious problems. The law making and jurisdictional sector is notoriously slow, and will have to make provisions for innovations that at the time of promulgation are not concrete in every detail. Problems with the regulation of the current internet prove how difficult this can be, but also, that putting the lid back on a box of a more or less anarchic sector can be still more challenging. The concept of indefinite legal terms within laws and regulations may offer a way out. Their concrete interpretation is typically left to Case Law as Quantum applications evolve, but still leaves unresolved the problem of lagging behind market developments. In a dialectic perspective, within the core businesses of all of the established industries mentioned in Chap 2, tight administrative standards and high frequency and volume reporting requirements are a global standard. Faster and even more tight and reliable control will be made possible through QC. From the perspective of politics and political science, shaping this legal and regulatory framework is just one requirement. From a broader perspective, the established concept of Digital Sovereignty must be rethought. QC is the first of the IT revolutions where with China a non western, non democratic state plays a prominent role on the leading edge of development in all aspects (hardware, software, application). As in all centralized societies, far fewer obstacles to implementation will be experienced along the realization path. Furthermore, the breaking of encryption methods has attracted interest of Intelligence agencies – around the globe and serves to funnel funds into development. In the Western World, hardware development is centered around the largest among multinational enterprises with the US government also taking substantial part. Europe has been lagging behind and given the speed and productivity of research will probably be unable to catch up. A strategy of "leapfrogging" may instead be advisable (probably also for the Global South): trying to be faster on the way to widely used, secure and socially accepted everyday applications, and having a nimble innovation system for the hundreds of smaller post-breakthrough innovations. To pave this avenue, a close and speedy collaboration between all societal stakeholders is called for.

---

[7] The problem of loading large amounts of data into Quantum storage may be seen a second obstacle.
[8] An eavesdropper will store an encrypted message in the hope that with the advent of code breaking techniques he may at a later point in time be able to decode it.

### 2.5.2 Security

For security production in cyberspace, fundamental changes will occur in the dimension of the threat situation (from individual companies to the entire economy) and the protection actions (post-quantum cryptography and quantum cryptography) as well as the analysis options for assessing threat situations. This makes security research a lot more technical - quantum technical. In the next few years, security researchers will no longer be able to avoid using quantum computers or including them as a central factor in their calculations and analyses (See also 2.5.1). This also applies in particular to the security-providing authorities. New questions will arise for security research that relate to the emergence of probabilistic algorithms as a side effect to the advantages that arise from the quantum computer in terms of speed and accuracy. Security research mss be able to assess the hazards that arise from quantum computing, but in the future it will also need to draw on the technology as a tool to focus on those developments that previously required too much computing power, such as calculating the superposition of sentiments in the population toward actions and fault lines, or analyzing emergent hazards just to name an example (NEASQ a, n.d.).

### 2.5.3 Energy

Energy is a critical sector with multiple relevant dimensions for social sciences and humanities. It is intrinsically connected to the utilities that allow everyday operations of any society. Energy security issues, such as vulnerabilities in critical infrastructure or access to regional or global energy markets, are important on national levels. Sustainability and cost of access to energy resources is an essential requirement for any industry. Furthermore, processes such as energy transition (*Energiewende*) require development of novel and innovative approaches compared to the established rules of the public energy infrastructure management strategies and industries. Considering all these, the risks and opportunities that quantum computers might present should be assessed and addressed. There will be plenty of room for research to be undertaken by humanities and social sciences.

### 2.5.4 High Performance Computing (HPC)

HPC resources are actively used in all fields of STEM research, however, they are costly both in terms of financial resources and carbon footprint (Allen 2022). Furthermore, supercomputers were not always the go-to tools of researchers and companies. Considerable effort went into making scientific computing a powerful tool and enabler of research. Now, with the oncoming of the quantum computing era, new opportunities arise for scientific computing. The promise of fully realized large scale fault-tolerant quantum computers that work hand in hand with classical supercomputers is to reduce both of these costs, and even further enable researchers to do novel research. This is, of course, not a straightforward step where research groups just re-write some parts of their simulations to make them run on these new devices. As mentioned above, quantum computers provide speed-ups for only a subset of algorithms, and even in this subset there are rank orders between the amount of speed-up quantum computing can provide. Therefore, much research and effort is needed to explore where a quantum assisted supercomputer can benefit research greatly and where it can't.

**2.5.5 Transportation**

Quantum computing can be considered as particularly relevant in this field for social sciences and humanities in two aspects. First, studies on public transportation resources and infrastructure. Second, studying the complex transportation systems themselves.

For the first point, as the development of autonomous vehicles and smart city technologies mature, distribution and management of public resources become more and more a data analysis problem. Through creating digital twins of cities and certain environments, researchers in mobility studies can test out many scenarios without the need for actually implementing them. These scenarios can be transformed into visual experiences and be used for stakeholder engagement.

Second, the mobility systems themselves are multi-faceted. There are already HPC enabled studies on these systems (Fraunhofer 2022). However, due to the increasing complexity with scale, running a comprehensive simulation even for a small city is computationally too demanding. In the current models there needs to be many simplifications and reductions of complexity for these systems to become simulatable. As large scale fault-tolerant quantum computers develop, certain parts of the complexity might be outsourced to QPUs in the HPC systems, enabling more detailed simulations that can provide further insights into the current and potential mobility and transportation systems.

**2.5.6 Health**

Health issues and medical technologies have been gaining popularity even before the emergence of COVID-19. With all the digital measures levied during the period of this pandemic, participation and involvement of social scientists to the processes became more important. Furthermore, there are proposals like the European Health Data Space (European Commission/Directorate Health Data Space n.d.) that is expected to generate huge amounts of data, which can enable all kinds of research on identifying the relations between socially relevant factors like education and socioeconomic status, and medical conditions. Quantum computers and quantum enabled applications can play an important role here, as they are expected to be used for: (i) diagnostic assistance, (ii) precision medicine, and (iii) pricing.

First, any researcher studying either the network of socio-technical relationships or the consequences of implementing these technologies would highly benefit from having a general understanding of what quantum computers can and can't do. A similar transition period has been happening for the A.I. studies, where initially researchers from non-technical fields had limited ideas on what A.I. systems can accommodate for, but as the topic matured, the non-technical researchers gained a general understanding of what A.I. systems are capable of.

Second, actually utilizing quantum computers for performing scientific studies in social sciences and humanities requires a level of expertise that, since no packaged software with GUI like VOSViewer or SPSS is expected to appear during the emerging years of quantum computing. There are many training programs and bootcamps on introduction to quantum programming and even for specialized fields like finance, but the tools available for social science research using quantum computers don't exist and a particular set of expertise is required to develop those, especially for the tools that deal with sensitive data such as health and medical records.

**2.5.7 Natural Language Processing**

In particular, the humanities and social sciences depend on the analysis of communication, on the analysis of natural language. There are no contracts, no political trades or laws without language and every discussion, every historical fact is transmitted through language. Therefore, it could be that this field holds one of the most important transformations for our sciences. There are countless examples that could be given to show the potential impact on our sciences that the use of technology will have. These include all the tasks of systematization, such as the construction of repositories or schema and network developments, but also the development of systematics based on existing data sets. The QNLP is relevant for all sciences that rely on language analysis, from history, law, political science, sociology, and religious studies to criminology and other "minor" specialties. QNLP changes the way of NLP as it "treats language as a quantum process and interprets sentences as circuits by using categorical quantum mechanics" (Garcia Molina 2021). Tools like Python and R seem to be applicable for qc, and a first library for Python to use for a quantum computer is getting developed and found at Github, yet it seems that QNLP applications are still waiting to be developed.

**3. Development**

At this point, it would be possible to work through the individual grants, policy networks, and research alliances that feature the specific applications of quantum-related research in each case. Instead, we consider it useful to make general comments that can more clearly elaborate the structures of Quantum Humanities Research in regard to development. Certainly, completely different approaches to methodology are possible here, because from observation of laboratory development and conversation to interviews, network analyses, policy analyses, or legislative history analyses, completely different research scenarios are conceivable that can encompass the field. However, development as a process of shaping technology within a society is not meant by development alone (Madrigal 2012). transformative politics as a purposeful shaping of a complete change is here described with Netdoms, Styles, Sentiments, Ideology and other emotion-associations. What is the person who builds this actually thinking? What is he driven by?Here, however, not only emotions are meant, but also a frame of action in which a person stands can be examined. From the laws, which pretty much everything, which surrounds these researchers, coined, is also their action of these normed. How do the researchers deal with this and what do they need?

In describing the implications that an application will have, we have already been quite speculative, because today, unfortunately, it is not yet possible to foresee what exactly the interference-free quantum computer will be capable of. The development as an anticipation has here certainly bonds to the futurology (Gerhold/Brandes 2021). The weighing of (future) interest positions could be important for this section. Who is interested in this topic and how will the citizen group of x be affected by this future technology offer and how can this consideration be integrated into the design of the technology? (Knappe, H., Renn 2022) Comparative studies on the National Research Systems of quantum computers, on the laboratory conditions of hardware or software production, and on the social disputes that do not exist yet or we are not aware of. Yet it is precisely these questions that are very exciting. At this point, we have to accept an honest blank and admit that we are still too early.

**4. Approaching a Research Program**

We have depicted the four fields shown-applications, implications, reflections, and development-as intertwined fields. We would like to briefly conceptualize this.

We have shown that a largely open theory, grounded in sociology, can be a basis for various social sciences and humanities when it comes to inferring the meaning of technology. The Organizational Shell suggests that we can use it to work on interrelated issues within a major transformation, and thus break down the transformative aspects into individual themes that are conducive to the big picture. We were able to define four distinct, temporally related and overlpping phases in which the processes for transformative change take place in different ways. For example, the application development phase is an intensely transformational phase that affects society in multiple ways and triggers transformative processes. Since quantum computing represents a radical leap innovation, we now understand that the development of a research framework can well replicate these leap innovations in their socio-institutional design. What we do not know, however, is whether the individual research phases we have defined in time, such as development, have an equal effect on the overarching representation of changes.

This is a first and preliminary review, it should help to get into the discussion and to evoke new considerations - also deepening considerations. Since funding for the humanities and social sciences from third party donors has been sparse so far, this is also a call for attention to develop new third party funding that will help the community to engage and thus also influence a responsible design of quantum technology. The distinction between application, implication and reflection has enabled us to think in a structured way about the transformation processes that QC has released. Emerging from the fields discussed here, what innovations does quantum humanities offer and how can this be meaningfully embedded in Humanities and social sciences?

Quantum humanities basically deals with a technology and uses it or reflects it in the fields of humanities and social sciences. We have fanned this out systematically on the basis of application, reflection and implication and development. However, the contribution presented here can be an example for a structural consideration of what exactly constitutes quantum humanities as a research field. First, it is a technology-related science, and second, it is a science that presents itself (with Geertz) as a cultural practice of the everyday handling of technology - be it costumes in film, the development of law, or the development of significant infrastructure, and so on. We have been able to demonstrate here that the innovations associated with the quantum computer will have a massive impact on the shaping of science and coexistence, to such an extent that it will not be a single science that is entrusted with technology assessment but an umbrella.

The very development shown here suggests that we should make the solution to the task as multivariate, as the task itself is. And here an important restriction must be made, because with quantum technologies a whole set of techniques is meant, which could not come up here at all (and this will be a future task for the development of a research program) because quantum sensing, quantum communications and quantum metrology have not been covered in this work and also the innovations in these areas can have a large effect on social science and humanities studies, which are not yet fully foreseeable today. A new emerging topic will certainly be the identification of narratives and metaphors in order to bring quantum knowledge to the public (Grinbaum 2017). Scientific work on quantum computing or scientific work about quantum computing require cross-scientific critique and reflection. What happens to the energy sector has an impact on humanities and social science. System knowledge, target knowledge, and transformation knowledge come into focus and contribute to an overview, being able to classify aspects. Problem-solving and decisions in one discipline become relevant to all disciplines. The internal self-organization of our science is thus (perhaps!) subject to change.

Fundamental insights into the basis of knowledge might be gained that will have a broad impact on the way we think, act and communicate. It will possibly change how we do something. Our contribution is an elaboration of the core themes to ensure a comprehensive mapping of the subject matter that grounds the research program and allows for critical engagement with and in the field. This serves as preliminary work for assigning quantum humanities a meaningful role within the humanities and the social sciences.

What have we learned about the quantum computer and its applications, and what processes of change will it trigger in society and what transformative paths will development take? The quantum computer will trigger a leap in some areas, although it can be seen that our previous computers are also already quite good (as for example in the calculations of quantum chemistry) and not leap innovations everywhere, but incremental innovations will happen. Our research program tends to put human actions in a temporal sequence context and relate socio-institutional rules to them. However, this also provides a qualitative approach to national innovation systems and their importance for societal transformation, which helps us to apply the political interventions in technology development and design to a societal-institutional meaning. We will see changes in finance, but also in the development of materials and surfaces that will become important, for example, in combating climate change or in a new type of medicine. It will be important to work out the specific regulatory needs for the individual innovations and there will certainly also have to be new powers of intervention and new concepts of protection. The research program presented here can well complete the individual work into the depth of change with an interconnection of the individual transformational objects such as law or standardization. We have already argued that the research program and its disaggregating impact has so far only been disaggregated in the area of application and more time needs to be spent on making the other temporal phases of activity the focus of elaborations, but we also need to make an increased effort in the area of individual institutions of society on how individual transformations affect different institutions of society and how the interplay can be better captured. On the one hand, new industries could emerge (manufacturing industry, user industry) but also legal frameworks could be adapted (property rights of water in the air) and new educational processes (water agriculture) could develop and a new settlement in areas that were previously hostile to living beings, such as deserts, could again trigger new processes.

Quantum humanities is an interdisciplinary field that explores the intersection between quantum computing and the humanities, including the social sciences and the arts. It aims to understand the impact of quantum computing on society and to consider the theoretical and philosophical implications of these technologies for the humanities. The applications of quantum computing that have the potential to significantly impact the humanities and social sciences include finance, security, energy, high performance computing/scientific computing, transportation, health, and natural language processing. Quantum security is a particularly important area, as it deals with the use of quantum algorithms to secure digital systems. In the field of finance, quantum computing has the potential to revolutionize investment and risk analysis, as well as to enable more efficient and accurate financial modeling. In the field of energy, quantum computing could be used to optimize the design and operation of renewable energy systems and to improve the efficiency of chemical reactions. In the field of transportation, quantum computing could be used to optimize logistics and supply chain management. In the field of health, quantum computing could be used to analyze and interpret medical data and to design personalized treatment plans. In the field of natural language processing, quantum computing could be used to improve the accuracy and efficiency of language processing tasks, such as translation and summarization. Overall, the advancements in quantum computing have the potential to significantly impact the way we live, work, and think, and they raise important questions and implications for the humanities and social sciences to consider. It is noteworthy that it was possible to show how far the new applications could have an impact on the further transformation of societies when the technology has matured further. And because this technology has great potential in terms of its disruptive capabilities, it is worth taking a deeper look into applications even where transformations do not have obvious effects on the humanities or social sciences. Of course, some of the points made here are still dreams of the future. Nevertheless, it will be exciting to take a close look at these developments. We would also like to point out the other fields of Quantum Humanities Research that are still to be worked out here, because now is precisely the time to take a closer look at developments and their social situatedness and to ask questions about the conditions of emergence of technical development in the quantum realm.